# DFT Studies of Indium Nanoclusters, Nanotubes and their Interaction with Molecular Hydrogen

A. Hussain[1,2*], M. W. Baig[3] and N. Mustafa[1]

[1]*TPD, Directorate of Science, PINSTECH, P. O. Nilore, Islamabad, Pakistan*

[2]*DNE, Pakistan Institute of Engineering and Applied Sciences, Nilore, Islamabad, Pakistan*

[3]*Polymer Chemistry Laboratory 40, Department of Chemistry, Quaid-i-Azam University, Islamabad, Pakistan*

*ahmohal@yahoo.com; mylifeischemistry.wasif@gmail.com; naeem_mustafa@hotmail.com*

A B S T R A C T

Density functional theory calculations have been performed on Indium nanoclusters ($In_n$, n= 3 to 10) to explore the relative stability among their different isomers and interaction with $H_2$. Geometry optimizations starting from initial candidate geometries were performed for each cluster size, so as to determine a few low energy isomers for each size. Clusters with planar configuration and high symmetry are found to be more stable. For n=8, there comes transition from 2D to 3D structures. Energetically favorable isomers of indium nanoclusters for each size were considered to get $H_2$ adsorbed. In general $H_2$ interaction with these clusters is weak but with those comprising of some odd number of atoms i.e. 5, 7 and 9 is considerable. Indium nanotube also indicates $H_2$ adsorption but $E_{ads}$ increases many folds on introduction of defect in the tube. On the basis of DFT investigations, it is suggested that apparently indium nanoclusters and tubes of specific size seem better candidates for materials to store hydrogen.

## 1. Introduction

Nano-clusters are collection of three to several hundred atoms [1]. The main reason of researcher's attention in nanoclusters is because of their potential uses in quantum computers & dots, chemical sensors, industrial lithography, applications with photochemical pattern and their use in light emitting diodes [2-3]. Abnormal energetic, thermodynamic and kinetic effects are governed by quantum confinement effects in nanoclusters [4]. Nanoclusters have been synthesized massively for last few decades by several groups around the globe due to their extraordinary catalytic and optical applications [5].

It has been made possible due to initiation of DFT and other computational techniques to model and predict the geometrical, structural, catalytic, optical etc. properties of metals and metal oxide nanoclusters [6]. Clusters of transition metals and their alloys have been widely studied using DFT. Investigation of P.E. surfaces for metals nanoclusters support us to optimize their structures [7]. Isomerism in metals' nanoclusters semiconductor along with their alloys have been broadly investigated by this theory as their ground state electronic energy is completely reliant on geometry of nanoclusters. Additionally, the geometry being the single parameter tells their specific optical properties & catalytic activities [8-9]. Especially, gold and silver nanoclusters have been computationally studied by numerous research groups [10-11]. DFT studies are rarely used for elements of p-block in research. Truhlar et al. have in recent times investigated nano-thermodynamics & the structure of Al nanoclusters by searching the structures with global minimal-energy along with several local minima of higher-energy. They found leading structures of nanoparticles and clusters are dependent on size of particles & temperature and to determine the characteristics of nanoparticles, their stability and dominant structures we must take account into electronic structure as well as statistical mechanics [12].

Indium oxide (InO) structures of small clusters and their electronic and/or vibrational characteristics were explored by Mukhopadhyay et al. by DFT based first-principle method. They reported that neutral clusters of Indium oxide will probably be unstable if ratio of metal to oxygen will be more than one [13]. Walsh & Woodle have explained isomerism in $(In_2O_3)_n$ and predicted geometries of stable and metastable structures of $In_2O_3$ clusters and have found that they have a tendency in the direction of denser, lower symmetry structures approaching the bulk system at extraordinarily small molecular masses [14]. Indium nanoclusters have been studied experimentally keeping in view the importance for their thermal stability [15], plastic compatibility [16] and for their melting temperature as function of their size [17]. Indium nanoclusters' optical properties have been explored [18] for their applications in quantum devices [19] and quantum dots [20]. In addition to indium nanoclusters, indium nanotubes have also been synthesized by prototype free solvothermal path at various temperatures [21] and direct thermal evaporation of an indium metal source in atmosphere [22]. DFT studies have been reported for electronic properties of indium phosphide nanotube [23], while DFT studies for indium nitride nanotube predict their stability for their successful artificial synthesis [24].

---

* Corresponding author

To the best of author's information no one has analyzed indium nanoclusters and nanotubes through DFT regarding investigations for geometric and catalytic properties and hydrogen storage applications. Here we explore isomerism in indium nanocluster and study their interaction with molecular hydrogen. Hydrogen is considered as zero emission fuel and there are great efforts throughout the globe to find materials that can be used for hydrogen storage [25]. In addition to In isomers, study has been extended to investigate $H_2$ interaction with In nanotubes, where it is predicted as good candidate for hydrogen storage.

## 2. Computational Details

VASP is a package that performs simulations based on initial method [26] and it was used to get an iteration based result for Kohn-Sham Eqs. in a plane-wave basis set. The plane waves with kinetic energy (K.E.) limit ≤ 400 eV were considered. As suggested by Perdew & Wang (PW91), we have calculated the exchange-correlation energy within the generalized gradient approximation (i.e. GGA) [27]. The ion-electron interaction for In and H atoms has been designated by a method designed by Blöchl called projector-augmented wave (PAW) method [28]. It is fundamentally a technique that combines computational ease of pseudo potential approach with the precision offered by all-electron method [29].

Geometry optimization for all clusters reported in this study, their interaction with $H_2$ and In nanotube were carried out employing cluster model approach. A tetragonal unit cell having dimensions $12\times12\times18$ Å$^3$ was used for cluster calculations. A minimum vacuum distance of 10 Å was ensured in all directions to avoid any interactions between the repeating units. Nevertheless, dimensions of the unit cell were made two fold (24×24×36) for nanotube calculations. The nanotube composed of 120 atoms has length of 22.07 Å and diameter of 16.28 Å. The computation was carried out at 3×3×1 k-points for indium clusters and for indium nanotube at gamma 1×1×1 k-points. Monkhorst-Pack method was used for automatic generation of these k points [30]. A first-order Methfessel-Paxton smearing-function with a width 0.1 eV was used to account for fractional occupancies [31]. Partial geometry optimizations were performed including the RMM-DIIS algorithm [32]. Geometry optimizations were stopped when all the forces were smaller than 0.05 eV/Å. The adsorbate-surface coupling was neglected and only the Hessian matrix of the adsorbate was calculated [33]. Using a 10x10x10 Å$^3$ cubic unit cell with the help of non-spin polarized computations, closed shell $H_2$ molecule is optimized at the point.

The spin-polarized computations were also performed for some (particularly having odd number of In atoms) clusters. It was examined that the total energy was enhanced using spin polarized calculations but the relative difference of energy among various isomers of a cluster was not changed.

The adsorption energy of the adsorbates was calculated using the following equation :

$E_{ads} = E_{system} - E_{slab} - E_{H2(gas)}$

where,

$E_{ads}$ = Adsorbed energy of $H_2$

$E_{system}$ = Total energy of the system when $H_2$ is adsorbed on the cluster/nanotube

$E_{slab}$ = Energy of the cluster/nanotube slab

$E_{H2(gas)}$ = Computed energy of $H_2$ in gas phase

## 3. Results

Indium (In) nano-clusters with sizes 3-10 In atoms are explored. Various isomers of these clusters have been examined to explore geometrical properties, stability and symmetry. Nevertheless, we did not report all of them but a few among the most stable ones from each category. We report also the interaction of these clusters with $H_2$. The most widely used method to calculate van der Waals interactions - based upon the addition of pairwise $C_6/R^6$ terms to the internuclear energy term - permits for highest degree of flexibility to select independently suitable description of electronic structure, on top of which a suitable dispersion correction has been done in the DFT energy (DFT-D approach). While this semi-empirical method is implemented in VASP, DFT-D method is not considered to be reliable for metallic clusters [34]. In particular for molecular adsorption on metallic clusters, this method has the tendency to over bind. We have therefore preferred a more systematic approach within GGA framework. $H_2$ adsorption on the perfect and defected nano-tube was calculated & significant change in binding strength was observed. We could not find any relevant previous investigations to compare our results.

$$In_n\ (n = 3 - 5)$$

The three relatively more stable structures for 3, 4 and 5 indium atoms clusters are shown in Fig. 1. For the cluster comprising of three atoms, the most stable geometry is associated with triangular structure having $D_{3h}$ point group. The structure optimizes not to perfect but almost equilateral triangular structure as can be seen from the geometry depicted in Fig. 1a. Interatomic bond lengths have been measured to be 2.92-2.98 Å making bond angles 59° to 60.9°. The linear construction possessing $D_{\infty h}$ point group is 0.20 eV less stable energetically relative to most stable configuration explained above. Interestingly, the two In atoms are symmetrically 3.06 Å away from middle atom with bond angle 180° (see Fig. 1b). The V shaped structure which possesses $C_{2v}$ point group is the least stable among the three reported isomers with relative de-stability of 0.36 eV. Two terminal atoms as shown in Fig. 1c are symmetrically 3.02 Å far off from the vertex atom where bond angle formed is 110.3°.

Various sites were taken into consideration to examine the H$_2$ adsorption on the triangular structure. No interaction of H$_2$ was found with the structure (E$_{ads}$= -0.01 eV), that can be demonstrated from its staying at a large distance of 4.84 Å from the cluster.

In case of In$_4$ (In isomers consisting of 4 atoms), three lowest energy structures are presented as shown in second row of Fig. 1. The square structure with $D_{4h}$ point group having bond lengths of 2.99 Å is observed to have lowest energy. The other two structures namely Y shape and T shape belonging to $C_{2v}$ and $C_{3v}$ point groups are 0.63 and 0.69 eV less stable, respectively; as compared to the square model. In T-shape model, the two atoms making almost linear construction are symmetrically 2.99 Å distant from the central atom while perpendicular atom is relatively closer as depicted in Figure 1e.

H$_2$ interaction with these structures is also negligibly small and consequently remains at large distance from the structure. Interaction energy is on the order of -0.01 eV.

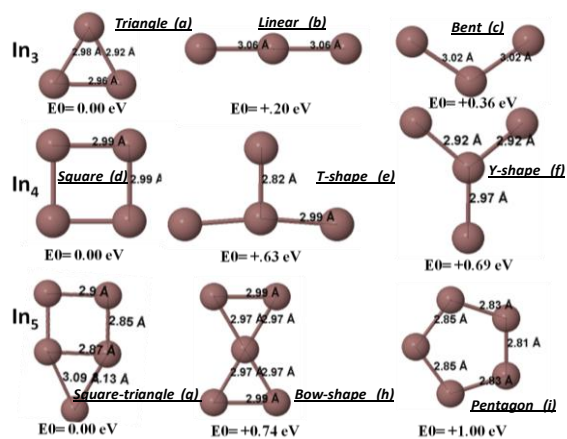

Fig. 1: Isomers of three, four and five atoms In clusters. Their relative stability and geometrical information is given in each case

The 3rd row of Fig. 1 shows In isomers consisting of 5 atoms. The lowest energy In$_5$ cluster consists of a square of 4 In atoms with the 5th In atom forming the In$_3$ triangle on one side of the In$_4$ square. This configuration possess $C_{2v}$ point group. As it can be seen in Fig. 1g, all bond lengths neither in square part nor in triangular portion are identical. While the other two structures are named as bow (comprising of two triangles in horizontal plane) connected via central atom) and pentagon structures. These isomers belong to point groups $D_{2h}$ & $D_{5h}$, and are 0.74 and 1.0eV, respectively destabilized relative to the most stable formation. The bow shape structure possesses bond lengths of 2.97 Å that connect all atoms to the central one while terminal atoms are 2.99 Å far off from each other. A planar pentagon conformation optimizes as shown in Fig. 1i where inter-atomic distances are observed in the range of 2.81 to 2.85 Å. The central atom of bow structure interacts with H$_2$ molecule with presentable adsorption energy of -0.21 eV. All structures where H$_2$ interacts remarkably are shown in Fig. 5. As depicted in Fig. 5a, the H$_2$ stays at a distance of 3.97 Å from central atom. Here H$_2$ is closer if compared with previous structures where interaction was negligible.

$$In_n (n = 6 – 8)$$

Isomers of Indium containing 6 atoms are shown in Fig. 2. The 2 adjacent squares (named fused squares) conformation with $D_{2h}$ point group having bond lengths of 2.90 Å except terminal sides - where inter-atomic distance is 2.95 Å - is of minimum energy (Fig. 2a). The other two structures named as big triangle (containing 4 small triangles) and incomplete hexagon belonging to point groups $D_{3h}$ and $C_{2v}$ are 0.29 and 0.34 eV energetically higher (less stable), respectively. The sides of the small triangles coincident with planar big triangle are 3.07 to 3.08 Å in length while inner sides measure 2.95 to 2.96 Å; see Fig. 2b. The incomplete hexagon structure possesses bond lengths of 2.87 to 3.14 Å. Here relatively less saturated In bonds are more stretched causing significant variation in inter-bond distances as shown in Fig. 2c. However, the corresponding distances on the two sides are equivalent. The H$_2$ interaction with this series of indium isomers is also negligibly small, the highest value of H$_2$ E$_{ads}$ being - 0.065 eV with incomplete hexagon.

For Indium nanoclusters comprising of seven atoms, the three lowest energy configurations are reported, as shown in 2nd row of Fig. 2. The hexagonal structure with $D_{6h}$ point group possessing inter-atomic distances of 3.03 – 3.09 Å is of minimum energy relative to the other two formations namely heptagon and incomplete heptagon. The incomplete heptagon is pentagon with two additional In atoms, which forms two triangles. These configurations belong to point groups $D_{7h}$ and $C_{2v}$ and, are 0.37 and 0.87 eV less stable, respectively; relative to the most stable hexagon configuration. A planar heptagon conformation as shown in Fig. 2e makes a circular ring of atoms where inter-atomic distances vary in the range of 3.14 to 3.18 Å. A cluster depicted in Fig. 2f – named incomplete heptagon – resembles to a shape formed as a combination of a bow and a pentagon configuration. However, there is a remarkable difference in the bond lengths joining different atoms as visible in Fig. 2f.

The most important feature of this isomer of In nanoclusters is its affinity for $H_2$ molecule. $H_2$ adsorbs with central atom having substantial adsorption energy. Consequently, $H_2$ optimizes at a shorter distance of 3.29 Å, relative to the other structures discussed above where $H_2$ does not interact noticeably. However, this result loses its importance because this interaction is not with the most stable configuration.

On addition of one more In atom i-e. for $In_8$, three lowest energy structures are reported as shown in 3$^{rd}$ row of Fig. 2. The cubic structure with $D_{4h}$ point group possessing all bond lengths of 3.09 Å represents the most stable configuration. The other two conformations namely bi-pyramidal hexagon and heptagon with point group symmetry of $D_{6h}$ and $D_{7h}$, respectively; possess energetically higher (less stable) value. Bi-pyramidal hexagon comparative to cubic configuration is 0.13eV less stable while heptagonal configuration is significantly 0.89eV higher energetically. In a non-planer bi-pyramidal configuration the six indium atoms forming the ring have symmetric inter-atomic distances of 2.81 Å. However, two atoms staying at mid-point are mutually separated by 4.12 Å. In a planar heptagonal structure, seven atoms lying in a ring are mutually 2.93 – 3.01 Å far-off but more distant 3.35 – 3.50 Å from the central atom. These configurations do not have any significant interaction with $H_2$.

### $In_n$ (n = 9)

For $In_9$ atoms nanocluster four lowest energy structures are presented in Fig. 3. The cubic structure which exhibits the structure of a body centered cubic metal is investigated as the conformation with the most stability among the reported 4 configurations. All the atoms at the corners of the cube possess corner to corner bond lengths between 3.40 – 3.44 Å while these are 2.97 Å away from the central atom. This configuration has $D_{4h}$ point group symmetry. If this central atom is shifted at the outside of a cube, the configuration adopts the shape of a cage as shown in Fig. 3b. This isomer is 0.25 eV less stable relative to the most stable configuration. The lower 4 atoms making the bottom of the cage are symmetrically 3.05 Å spaced. However, the 4 atoms making the 2nd layer get wider being 3.50 – 3.52 Å spaced. The height of this layer relative to bottom layer is 2.92 Å. The ninth top covering atom is symmetrically 3.03 Å away from 2nd layer atoms.

The hexagonal bipyramidal structure has $D_{6h}$ point group. The atoms perpendicular to the plane have axial bond lengths equal to 2.99 Å while equatorial bond lengths equal to 3.09 Å. This configuration is 0.43eV destabilized than the most stable structure presented in this series of 9 atoms. The least stable isomer is crown shaped with $D_{4d}$ point group presented in Fig. 3d. This crown shape configuration is 0.51 eV destabilized energetically than the most stable formation. Four indium atoms form the base in a square geometry; the next 4 atoms also make square shape (each atom at the mid-point of two adjacent base atoms) but significantly shrink relative to base atoms, while the last atom occupies the top location. Mutual distance among various atoms can be seen in the Fig. 3d. The cubic structure being most stable among the 4 reported $In_9$ isomers interacts with $H_2$ with an $E_{ads}$ of - 0.18 eV. The $H_2$ molecule as shown in Figure 5(b) stays closer (2.95 Å) to indium atom at one corner of the cube in optimized position.

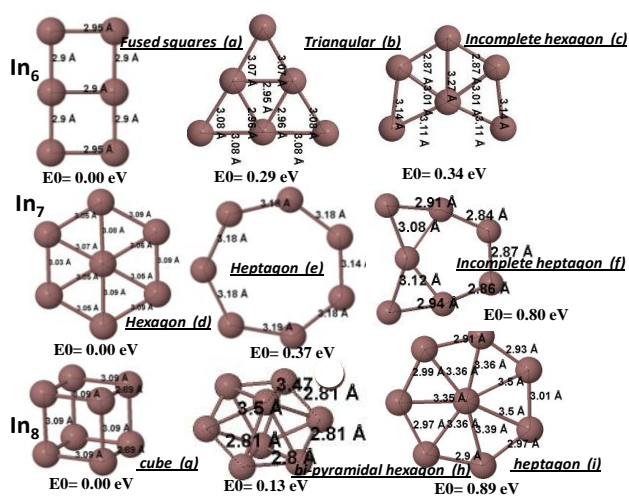

Fig. 2: Isomers of six, seven and eight atoms In clusters. Their relative stability and geometrical information is given in each case

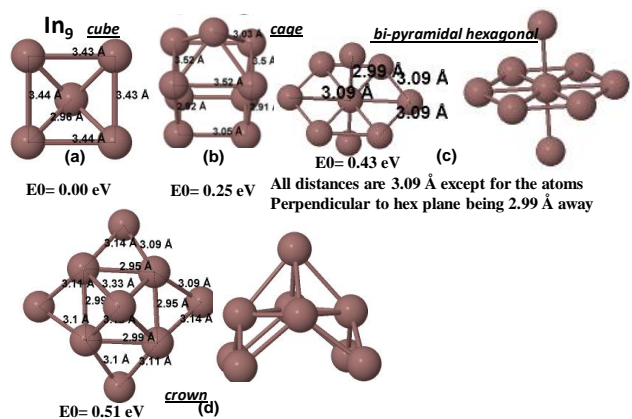

Fig. 3: Isomers of nine atoms In clusters. Their relative stability and geometrical information is given in each case

We did Bader charge analysis [35] to examine the charge transfer among the In atoms in a cluster as well as between In atoms and H-atoms. Redistribution of charge among different In atoms occurs at large scale. This charge transfer is intensified when $H_2$ interacts with an In atom of a cluster. For example, for the structure namely bow shape (consisting of 2 triangles) shown in Figure 5a the valance charge on different atoms varies between 2.86 e and 3.43 e. But upon physisorption of $H_2$, the valance charge contained by an In atom is as low as 1.62 e and as high as 6.08 e. However, the charge transfer is less in bigger clusters e.g. 1.77 e to 4.94 e in case of cubic and 2.89 e to 3.35 e for cage like structure of In isomers consisting of 9 atoms.

When $H_2$ interacts with a cluster, one of the H-atom is always oxidized (maximum up to 0.94 e) and other is reduced maximum up to 1.05 e. However, overall $H_2$ molecule is reduced; the highest value being 0.03 e. This charge gained is responsible for the bond length enlargement (0.76 Å) compared to gas phase computation of 0.74 Å. This small bond activation is result of additional charge received in the anti-bonding molecular orbital of $H_2$.

$$In_n (n = 10)$$

We have optimized 4 different isomers of 10 In atoms as shown in Fig. 4. The configuration consisting of cube with two additional In atoms lying at the opposite sides of the cube constructing 4 triangular geometries on each side depicted in part (a) of Fig. 4, forms the most favorable geometry. This configuration belongs to $D_{4h}$ point group. Mutual interatomic distances can be seen in the Fig. 4a. The slightly low stable (+0.15 eV) is the diamond structure like conformation with $D_{4d}$ point group. Different atoms are 2.96 and 3.06 Å far off mutually as shown in Fig. 4b. The two central atoms forming axis are 3.19 Å from each other. The third and last structure is referred as the staggered pentagons possessing $D_{4d}$ point group and 0.67 eV energetically destabilized relative to the most stable above mentioned configuration. The staggered pentagons is a construction where indium atoms' bond lengths in a plane are 3.0 – 3.03 Å but triangular distances lying between the two pentagons are 3.13 and 3.16 Å. This category of In isomers does not exhibit any remarkable affinity for $H_2$ as far as reported structures are concerned.

### 3.1. Indium Nanotube ($In_{120}$ and $In_{119}$)

Indium nanotubes comprising 120 atoms with 16.38 Å diameter were investigated using DFT. Indium nanotube consists of hexagons that are wrapped around imaginary cylinder to give tube like structure where all the atoms are equally coordinated as shown in Fig. 5. Thus an analogous to carbon nanotubes structure is constructed of indium nanotube. However, here interatomic distance between the coordinating atoms is uniform (2.86 Å) Interesting thing with In nanotube is its interaction with hydrogen molecule. It is found that hydrogen molecule physisorb on either outside or inside of indium nanotube releasing same amount of $E_{ads}$ i.e. -0.12 eV. Hydrogen stays at a distance of 3.44 Å. Defect was introduced in indium nanotube and its ground state energy was computed. $H_2$ molecule was placed near the defect site. Consequently, a substantial rise in $E_{ads}$ of $H_2$ on introduction of defect was noted. The $E_{ads}$ increased from -0.12 eV to -0.46 eV. As a result of this strong interaction, $H_2$ moves closer to the nanotube being stabilized at an equilibrium distance of 2.91 Å outside the tube. Similar calculations were performed by placing $H_2$ inside the tube. Similar adsorption energy was obtained; however, the molecule was slightly relatively nearer (2.79Å away from the nearest In atom). This $H_2$ adsorption on nanotube is depicted in Fig. 5 (c, d).

### 4. Discussion

From DFT studies of indium nanoclusters it can be concluded that they preferably optimize to form planar structures unless the number of indium atoms in a cluster exceeds 7. Because of inert pair effect of indium atom, significant energy gap between 5s & 5p orbitals exists caused by underlying filled orbitals 4d. Therefore, indium atom make use of its most outer p- orbitals for making clusters without going through any type of certain hybridization. This requires empty p orbitals in addition to σ-bonds for forming cyclic π-orbitals. These overlapped empty p-orbitals on distinct atoms can best fit up in planar structures giving up cyclic molecular orbitals that could be considered as holes. Lowest energy probable isomer is of $In_3$ cyclic triangular structure. Besides planarity shape, $In_3$ has high symmetry with that of $D_{3h}$ point group. This is one of the utmost significant features associated to stability of indium (In) nanoclusters. $H_2$ molecule has weak interaction as it interacts with extended empty p-orbitals on specific In atoms. Cyclic square structure of Indium $In_4$ has minimal-energy configuration. Besides planarity, $In_4$ has high symmetry with that of $D_{4h}$ point group. $In_5$, consisting of a square and an adjacent triangle has the most stable configuration and its higher stability on cyclic pentagon is vindicated by the point that pentagon structure requires indium atoms to show *sp* hybridization that is energetically costly. Therefore it is suitable to assume fused square triangle structure that is build up on use of just p-orbitals only. $H_2$ has some interaction with low coordinated atom at the periphery of bow shape isomer (at a distance of 3.57 Å). Peripheral atoms act together intensely with their empty p-orbitals with $H_2$ molecule and provides notable adsorption energy. The reported isomers of $In_6$ are highly symmetric and planar. The two adjacent squares namely Fused squares isomers are the very stable isomers and their stability can be illuminated on the same basis that it can be completely built on just valence p-orbitals. Big triangle and incomplete hexagon are very stable as well because of the planarity geometry. For $In_7$, three examined structures are hexagon, heptagon and incomplete heptagon. Hexagon is the most stable isomer and its coordination of all atoms and planarity is responsible for its stability. In complete heptagon likes to have hydrogen molecule at a minimum distance of 3.29 Å from central atom with adsorption energy -0.65 eV (highest one computed in this study). Relatively, this stronger interaction causes $H_2$ bond to activate up to 0.76 Å if compared to its gas phase computed bond length of 0.74 Å. However, this situation is hardly to occur because incomplete heptagon structure configuration is significantly (0.80 eV) destabilized than the most stable hexagon formation. For isomers comprising of 8 In atoms, we found conversion from 2-dimensional structures to 3-dimensional structures. Cubic structure, having 3D geometry, is highly stable composition with high symmetric $D_{4h}$ point group. It does not have planar construction however made by piling of two planar squares one above the other. Other reason that accounts for the extra stability might be due to maximum coordination of all atoms. For $In_9$, cubic geometry with 9th In atom at the center of cube among the 4 reported geometries (cubic, cage like, bipyramidal and crown like) is the most stable formation possessing high symmetry point group $D_{4h}$. This cubic make is the most stable owing to stacking planar squares just one above the other and central atom coordinating with corner atoms. In the category of 10 In atoms, the 4 reported relatively the most stable configurations form nonplaner (3D) geometries. Cube with 2 additional In atoms making triangular shape on opposite sides of the cube, configuration is the most stable structure whose stability is clear by the structure it has and that can be conveniently as well as completely made by use of valence p-orbitals. The diamond build up is slightly less stable and staggered pentagon is even more destabilized compared to the most stable cubic arrangement. Stacked pentagons are least stable because it requires less contribution of s-orbital i.e. $sp^2$ hybridization. $H_2$ interaction with these configurations is very limited.

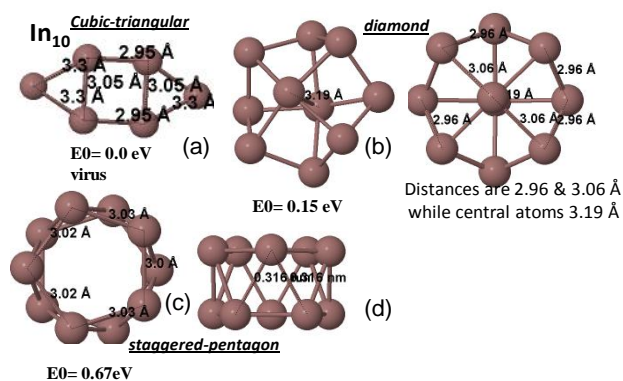

Fig. 4: Isomers of nine atoms In clusters. Their relative stability and geometrical information is given in each case

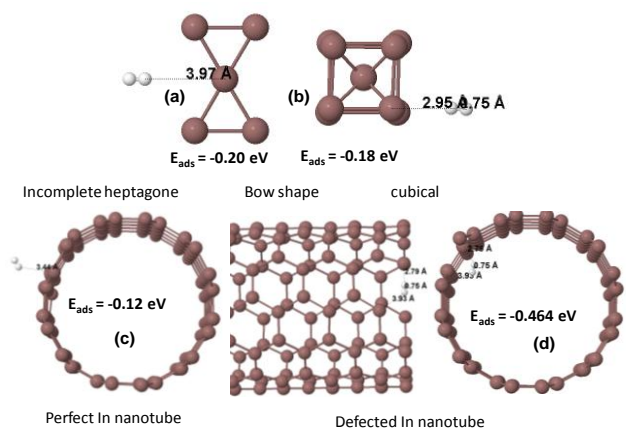

Fig. 5: The $H_2$ adsorption on In clusters and nanotube is presented. The geometrical information and $H_2$ adsorption energy is given in each case

Structure of indium nanotube consists of hexagons. It is the best appropriate unit cell for indium nanotubes as research on indium nanoclusters using density functional theory reveals that nanoclusters with planar configuration are more stable. It also reveals that clusters in which atoms coordinate to maximum level have higher stability than the others. So hexagonal structure can wrap in the best possible way around a cylinder in which all atoms are equally coordinated except terminal ones. Since large numbers of atoms are involved to give a stable tube like structure, thus to gain stability, promotion of an electron from low lying 5s orbital takes place in such a way that each atom is in $sp^2$ hybridized configuration. Empty p-orbitals make empty π-clouds around the whole nanotube. These are π-clouds that can be treated as sea of holes. In indium nanotube empty π-molecular orbitals which behave as sea of holes to symmetric molecule like hydrogen and develop attractive interaction with these deficient π-molecular orbital. And these interactions of hydrogen molecule with this sea of holes of indium nanotube results in adsorption of $H_2$. Introduction of defect in In NT by taking away an In atom, generates more holes and the value of adsorption energy calculated was enhanced substantially, it supports the truth that symmetric molecules like $H_2$ have affinity for holes and if the $H_2$ molecule is just in front of the defect introduced in the tube then the most stable configuration is attained.

## 5. Conclusion

DFT (Density Functional Theory) simulations are carried out for indium In nanoclusters of different sizes and shapes to ascertain the stability, geometrical information and $H_2$ adsorption. The results show that indium nanoclusters with high symmetry and planarity geometry are more stable. Additionally, the configurations which can be built by use of the clusters that require sp hybridization are less stable than those with only valence p-orbitals. There is a transition from 2D to 3D structures when number of In atoms in a cluster get to 8. Those 3-dimensional structures made by piling of planar units are more stable. The odd atoms cluster show superior interaction towards $H_2$ adsorption. Particularly; the clusters consisting of 5, 7 and 9 In atoms show high tendency towards $H_2$ adsorption. Indium nanotube in general has ability to interact with $H_2$ weakly but when defect is introduced in the structure, the hydrogen molecule interacts with significant adsorption energy. A slight activation in bond of $H_2$ also occurs. Hence, In clusters and nanotubes in defected form appear to be favorable materials for $H_2$ storage.